\documentclass[prl,superscriptaddress,unsortedaddress,twocolumn,showpacs,preprintnumbers,amsmath,amssymb]{revtex4}
\usepackage[dvips]{graphicx}
\usepackage{dcolumn}
\usepackage{amsmath}
\usepackage{times}

\begin{document}


\title{Accumulation of three-body resonances above two-body thresholds }

\author{Z.\ Papp}
\affiliation{ Department of Physics and Astronomy, 
California State University, Long Beach, California 90840 }

\author{J.\ Darai}
\affiliation{ Institute of Experimental Physics, University of Debrecen, 
Debrecen,  Hungary}

\author{J.~Zs.\ Mezei}
\affiliation{Institute of Nuclear Research of the Hungarian Academy of Sciences,
Debrecen,  Hungary}

\author{Z.~T.\ Hlousek}
\affiliation{ Department of Physics and Astronomy, 
California State University, Long Beach, California 90840 }

\author{C-.Y.\ Hu}
\affiliation{ Department of Physics and Astronomy, 
California State University, Long Beach, California 90840 }

\date{\today}

\begin{abstract}\noindent
We calculate resonances in three-body systems with attractive Coulomb
potentials by solving the homogeneous Faddeev-Merkuriev integral equations for
complex energies. The equations are solved by using the Coulomb-Sturmian
separable expansion approach. This approach provides an exact treatment of the
threshold behavior of the three-body Coulombic systems.
We considered the negative positronium ion and, besides locating 
all the previously know $S$-wave resonances, we found 
a whole bunch of new resonances accumulated just slightly above 
the two-body thresholds. The way they
accumulate indicates that probably there are infinitely many resonances 
just above the two-body thresholds, and this might be a general property of 
three-body systems with attractive Coulomb potentials.
\end{abstract}

\pacs{34.10.+x, 31.15.-p, 02.30.Rz}

\maketitle

The most common method for calculating resonant states in quantum
mechanical systems is the one based on the complex rotation of
coordinates.
The complex rotation turns the resonant behavior of the 
wave function into a bound-state-like asymptotic
behavior. Then, standard bound-state methods become applicable also for 
calculating resonances. The complex rotation of the coordinates 
does not change the discrete spectrum, 
the branch cut, which corresponds to scattering states, however,
is rotated down onto the complex energy plane, and as a consequence, 
resonant states from the unphysical sheet become accessible. 
By changing the rotation angle
the points corresponding to the continuum move, while those corresponding 
to discrete states, like bound and resonant states, stay.
This way one can determine resonance parameters. In three-body systems
there are several branch cuts associated with two-body thresholds.

In practice, the complex rotational technique is combined with some
variational approach. This results in
a discretization of the rotated continuum. 
The points of the discretized continuum scatter around the rotated-down 
straight line. So, especially around thresholds it is not easy 
to decide whether a point is a resonance point or it belongs 
to the rotated continuum. Moreover, variational methods approach states from
above, so resonances slightly above the thresholds may easily
get lost.

Recently, we have developed a method  for calculating resonances in 
three-body Coulombic systems by solving homogeneous 
Faddeev-Merkuriev integral equations \cite{fm-book}
using the Coulomb-Sturmian separable expansion approach \cite{pdh1}.
As a test case, we calculated the resonances of the negative positronium
ion. This system has been extensively studied in the past two decades 
and thus serves as test example for new methods.
We found all the $12$ S-wave resonances presented in Ref.\ \cite{ho} 
and observed good agreements in all cases.

We also observed that in case of attractive 
Coulomb interactions the Faddeev-Merkuriev integral equations 
may produce spurious resonances \cite{pla}, which are related to 
the somewhat arbitrary splitting of 
the potential in the three-body configuration space into short-range
and long-range terms. We could single them out by changing those parameters.
We succeeded in locating $10$ more
resonances in the same energy region, all of them are very close to the 
thresholds. These new resonances were published in Ref.\ \cite{pla}.

As our skill in applying our method developed we located more an more
new resonances just slightly above the two-body thresholds. They are all
aligned along a line in the complex energy plane pointing toward the thresholds.
It seems that there are infinitely many resonances accumulating at
the two-body thresholds.
Since our method is relatively new 
we briefly outline the basic concepts and the numerical techniques, 
specialized to the $e^- e^- e^+$ system 
(further details are in Refs.~\cite{pdh1,pla}).

The Hamiltonian of a three-body atomic system is given by
\begin{equation}
H=H^0 + v_1^C+ v_2^C + v_3^C,
\label{H}
\end{equation}
where $H^0$ is the three-body kinetic energy
operator and $v_\alpha^C$ denotes the
Coulomb potential in the subsystem $\alpha$, with $\alpha=1,2,3$.
We use throughout the usual configuration-space Jacobi coordinates
$x_\alpha$  and $y_\alpha$, where $x_\alpha$ is the coordinate of the
$(\beta,\gamma)$ pair and $y_\alpha$ connects the center of mass of
$(\beta,\gamma)$ to the particle $\alpha$, respectively. 
Thus  $v_\alpha^C$, the potential between
particles $\beta$ and $\gamma$, depends on $x_\alpha$.

The  Hamiltonian (\ref{H}) is defined in the three-body 
Hilbert space. The three-body kinetic energy, when the center-of-mass
motion is separated, is given by
\begin{equation}
H^0=h^0_{x_\alpha}+h^0_{y_\alpha}=h^0_{x_\beta}+h^0_{y_\beta}=
h^0_{x_\gamma}+h^0_{y_\gamma},
\end{equation}
where $h^0$ is the two-body kinetic energy.
The two-body potential operators are formally
embedded in the three-body Hilbert space
$v^C = v^C (x) {\bf 1}_{y}$,
where ${\bf 1}_{y}$ is a unit operator in the two-body Hilbert space
associated with the $y$ coordinate.

\begin{figure}
\resizebox{8cm}{!}{
\includegraphics{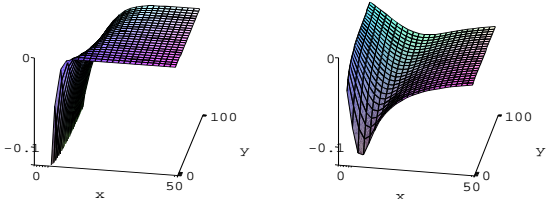}}
\caption{ $v^{(s)}$ and $v^{(s)}$ for an attractive Coulomb
potential.}
\label{vsl}
\end{figure}

In Merkuriev's approach to the three-body Coulomb problem
\cite{fm-book} the Coulomb interaction is split, in 
the three-body configuration space, into
short- and long-range terms 
\begin{equation}
v_\alpha^C =v_\alpha^{(s)} +v_\alpha^{(l)} ,
\label{pot}
\end{equation}
where the short- and long-range parts are defined via a splitting
function:
\begin{eqnarray}
v_\alpha^{(s)} & = & v_\alpha^C \;\zeta (x_\alpha,y_\alpha) \\
v_\alpha^{(l)} & = & v_\alpha^C \;\left[ 1- \zeta (x_\alpha,y_\alpha)\right].
\label{potsl}
\end{eqnarray}
The splitting function $\zeta$
is defined such that
\begin{equation}
\lim_{x,y \to \infty} \zeta (x,y) =
\left\{ 
\begin{array}{ll}
1, &  \mbox{if}\  |x| <  x_0 ( 1 +|y|/ y_0)^{1/\nu}, \\
0,  & \mbox{otherwise,}
\end{array}  \right.
\end{equation}
where $x_0, y_0 >0$ and $\nu > 2$.
So, in the region of three-body configuration space where
particles $\beta$ and $\gamma$ are close to each other
$v_\alpha^{(s)} \sim v_\alpha^C$  and $v_\alpha^{(l)} \sim 0$, otherwise 
$v_\alpha^{(l)} \sim v_\alpha^C$ and $v_\alpha^{(s)} \sim 0$. 
Usually the functional form
\begin{equation}
\zeta (x,y) =  2/\left\{1+ 
\exp \left[ {(x/x_0)^\nu}/{(1+y/y_0)} \right] \right\},
\label{oma1}
\end{equation}
is used. Typical picture for $v^{(s)}$ and $v^{(l)}$ are seen in Fig.\
\ref{vsl}.

In atomic three-particle systems the sign of the charge is always identical
for two particles. Let us denote those two particles
by $1$ and $2$, and the third one by $3$.
In this case  $v_3^C$ is a repulsive Coulomb
potential which does not support two-body bound states. Therefore the entire
$v_3^C$ can be considered as long-range potential and
the Hamiltonian can formally
be written in a form which looks like an usual
three-body Hamiltonian with two short-range potentials
\begin{equation}
H = H^{(l)} + v_1^{(s)}+  v_2^{(s)},
\label{hll}
\end{equation}
where the long-range Hamiltonian is defined as
\begin{equation}
H^{(l)} = H^0 + v_1^{(l)}+ v_2^{(l)}+ v_3^C.
\end{equation}
Then, the Faddeev method is applicable and, in this particular case, 
results in a splitting of the wave function into two components
\begin{equation}
|\Psi \rangle = |\psi_1 \rangle +
|\psi_2 \rangle.
\end{equation}
The components are defined by 
$|\psi_\alpha \rangle = G^{(l)} (z) v_\alpha^{(s)} |\Psi \rangle$,
where $\alpha=1,2$ and $G^{(l)} (z) =(z-H^{(l)})^{-1}$, $z$ is a complex number.

In the cases of bound and resonant states the wave-function components satisfy 
the homogeneous two-component Faddeev-Merkuriev integral equations
\begin{eqnarray}
 |\psi_1 \rangle &=& G_1^{(l)} (z)
v^{(s)}_1 |\psi_2 \rangle 
\label{fn-eq1} \\
 |\psi_2 \rangle &=& G_2^{(l)} (z)
v^{(s)}_2  |\psi_1 \rangle 
\label{fn-eq2} 
\end{eqnarray}
at real and complex energies, respectively.
Here $G^{(l)}_\alpha$ is the resolvent of the channel 
long-ranged Hamiltonian $G^{(l)}_\alpha(z)=(z-H^{(l)}_\alpha)^{-1}$,
where $H^{(l)}_\alpha = H^{(l)} + v_\alpha^{(s)}$.

Further simplification can be achieved if we take into account that 
particles $1$ and $2$ are identical and indistinguishable. 
Then, the Faddeev components 
$| \psi_1 \rangle$ and $| \psi_2 \rangle$, in their own natural Jacobi
coordinates, have the same functional forms
$\langle x_1 y_1 | \psi_1 \rangle = \langle x_2 y_2 | \psi_2 \rangle$.
On the other hand
$| \psi_2 \rangle = p {\mathcal P}| \psi_1 \rangle$,
where ${\mathcal P}$ is the operator for the permutation of indexes
$1$ and $2$ and $p=\pm 1$ denotes the eigenvalue of ${\mathcal P}$.
Therefore we can  determine $| \psi_1 \rangle$ from the first equation
only 
\begin{equation} \label{fmp}
| \psi_1 \rangle =  G_1^{(l)} v_1^{(s)} p {\mathcal P} | \psi_1 \rangle.
\end{equation}
It should be noted, that so far we did not make any approximation,
and although this integral equation 
has only one component, yet it gives full account both
of asymptotic and symmetry properties of the system.

We solve Eq.\ (\ref{fmp})
by using the Coulomb--Sturmian separable expansion approach.
The Coulomb-Sturmian (CS) functions are defined by
\begin{equation}
\langle r|n l \rangle =\left[ \frac {n!} {(n+2l+1)!} \right]^{1/2}
(2br)^{l+1} \exp(-b r) L_n^{2l+1}(2b r),  \label{basisr}
\end{equation}
$n$ and $l$ are the radial and
orbital angular momentum quantum numbers, respectively, and $b$ is the size
parameter of the basis.
The CS functions $\{ |n l \rangle \}$
form a biorthonormal
discrete basis in the radial two-body Hilbert space; the biorthogonal
partner is defined  by $\langle r |\widetilde{n l}\rangle=
\langle r |{n l}\rangle/r$. 
Since the three-body Hilbert space is a direct product of two-body
Hilbert spaces an appropriate basis
can be defined as the
angular momentum coupled direct product of the two-body bases 
\begin{equation}
| n \nu  l \lambda \rangle_1 =
 | n  l \rangle_1 \otimes |
\nu \lambda \rangle_1, \ \ \ \ (n,\nu=0,1,2,\ldots),
\label{cs3}
\end{equation}
where $| n  l \rangle_1$ and $|\nu \lambda \rangle_1$ are associated
with the coordinates $x_1$ and $y_1$, respectively.
With this basis the completeness relation
takes the form 
\begin{equation}
{\bf 1} =\lim\limits_{N\to\infty} \sum_{n,\nu=0}^N |
 \widetilde{n \nu l \lambda} \rangle_1 \;\mbox{}_1\langle
{n \nu l \lambda} | =
\lim\limits_{N\to\infty} {\bf 1}^{N}_1.
\end{equation}
Similar bases can be constructed for fragmentations $2$ and
$3$ as well.

We make the following approximation on Eq.\ (\ref{fmp}) 
\begin{equation} \label{fmpa}
| \psi_1 \rangle =  G_1^{(l)}
{\bf 1}^{N}_1 v_1^{(s)} p {\mathcal P} {\bf 1}^{N}_1
| \psi_1 \rangle,
\end{equation}
i.e.\ the operator 
$v_1^{(s)} p {\mathcal P}$ in the three-body
Hilbert space is approximated by a separable form, viz.
\begin{eqnarray}
v_1^{(s)}p {\mathcal P} &  = &
\lim_{N\to\infty} {\bf 1}^{N}_1 v_1^{(s)} p {\mathcal P}  {\bf 1}^{N}_1
\approx   {\bf 1}^{N}_1 v_1^{(s)} p {\mathcal P} {\bf 1}^{N}_1 
\nonumber \\ 
&  \approx &  \sum_{n,\nu ,n^{\prime },
\nu ^{\prime }=0}^N|\widetilde{n\nu l \lambda}\rangle_1 \;
\underline{v}_1^{(s)}
\;\mbox{}_1 \langle \widetilde{n^{\prime}
\nu ^{\prime} l^{\prime} \lambda^{\prime}}|,  \label{sepfep}
\end{eqnarray}
where $\underline{v}_1^{(s)}=\mbox{}_1 \langle n\nu l \lambda|
v_1^{(s)} p {\mathcal P}  
|n^{\prime }\nu ^{\prime} l^{\prime} \lambda^{\prime}\rangle_1$.
Utilizing the properties of the exchange operator ${\mathcal P}$
these matrix elements can be written in the form $\underline{v}_1^{(s)}= 
p\times (-)^{l^{\prime}} \; \mbox{}_1 \langle n\nu l \lambda| 
v_1^{(s)}|n^{\prime }\nu ^{\prime} l^{\prime} \lambda^{\prime}\rangle_2$.

With this approximation, solving Eq.\ (\ref{fmp})
turns into solving the matrix equation  
\begin{equation}
 \{ [ \underline{G}^{(l)}_1(z)]^{-1} - \underline{v}^{(s)}_1 \} 
\underline{\psi}_1 =0
\end{equation}
for the component vector
$\underline{\psi}_1 =
 \mbox{}_1 \langle \widetilde{ n\nu l \lambda} | \psi_1  \rangle$,
where  $\underline{G}_1^{(l)}=\mbox{}_1 \langle \widetilde{
n\nu l\lambda} |G_1^{(l)}|\widetilde{n'\nu' l' \lambda' }\rangle_1$. 
A unique solution exists if
and only if
\begin{equation}
D(z)\equiv 
\det \{ [ \underline{G}^{(l)}_1 (z)]^{-1} - \underline{v}^{(s)}_1 \} =0.
\label{fdet}
\end{equation}
So, to calculate resonances, we need to search for the zeros of determinant
$D(z)$ on the complex energy plane.

The Green's operator  $\underline{G}_1^{(l)}$  
is related to the Hamiltonian $H_1^{(l)}$,
which is still a three-body Coulomb
Hamiltonian and seems to be as complicated as $H$ itself.
However this is not the case. 
The only possible two-body asymptotic configuration  for $H_1^{(l)}$
is when particles $2$ and $3$ form a bound states and 
particle $1$ is at infinity. 
The corresponding asymptotic Hamiltonian is
\begin{equation} \label{htilde}
\widetilde{H}_1 = H^{0}+v_1^C.
\end{equation}
Therefore, in the spirit of the three-potential formalism \cite{phhky},
$\underline{G}_1^{(l)}$ can be linked to the matrix elements of
$\widetilde{G}_1(z)=(z-\widetilde{H}_1)^{-1}$ 
via solution of a Lippmann-Schwinger equation,
\begin{equation}
(\underline{G}^{(l)}_1)^{-1}= 
(\underline{\widetilde{G}}_1)^{-1} -
\underline{U}_1,
\label{gleq}
\end{equation}
where 
${\underline{\widetilde{G}}_1}  =
 \mbox{}_1\langle \widetilde{n \nu l \lambda} | 
 \widetilde{G}_1 |
 \widetilde{ n^{\prime}\nu^{\prime}l^{\prime}{\lambda}^{\prime}} \rangle_1 $
and ${\underline{U}_1} =
 \mbox{}_1\langle n\nu l \lambda | (v_2^{(l)}+v_3^C) | n^{\prime}\nu^{\prime}
l^{\prime}{\lambda}^{\prime}\rangle_1$.

\begin{figure}
\resizebox{8cm}{!}{
\includegraphics{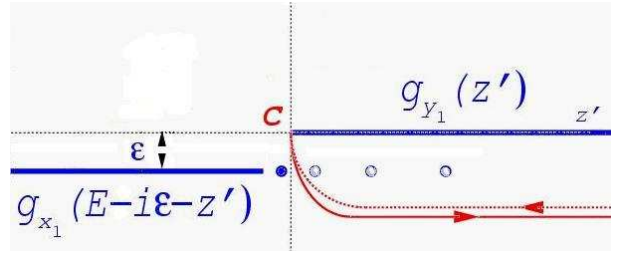}}
\caption{Analytic structure of 
$g_{x_1}(E+{\mathrm{i}}\varepsilon-z')\;g_{y_1}(z')$ 
as a function of $z'$, $\varepsilon=-\Gamma/2$. 
The Green's operator
$g_{y_1}(z')$ has a branch-cut on the $[0,\infty)$ interval, while
$g_{x_1}(E+{\mathrm{i}}\varepsilon -z')$ has a branch-cut on the 
$(-\infty,E+{\mathrm{i}}\varepsilon]$ interval and infinitely many 
poles accumulated at $E+{\mathrm{i}}\varepsilon$ (denoted by dots).
The contour $C$ encircles the branch-cut of
$g_{y_1}$ such that a part of it goes on the unphysical
Riemann-sheet of $g_{y_1}$ (drawn by broken line) and the other part detoured
away from the cut. The branch-cut and some poles of $g_{x_1}$ (denoted by full
dots) are lying on the physical Riemann-sheet, some other poles 
(denoted by empty dots) are lying on the
un-physical Riemann-sheet of $g_{y_1}$, respectively. 
Yet, the contour avoids the singularities of 
$g_{x_1}$.}
\label{contour}
\end{figure}

Now, what is remained is the
 calculation of the matrix elements
$\underline{\widetilde{G}}_1$, since the  potential
matrix elements $\underline{v}^{(s)}_{1}$ and
$\underline{U}_1$ can always be evaluated numerically.
The Green's operator $\widetilde{G}_1$
is a resolvent of the sum of two commuting Hamiltonians,
$\widetilde{H}_1 = h_{x_1}+h_{y_1}$,
where $h_{x_1}=h^0_{x_1}+v_1^C(x_1)$ and
$h_{y_1}=h^0_{y_1}$,
which act in different two-body Hilbert spaces.
Thus, $\underline{\widetilde{G}}_1$ can be given by 
a convolution integral of two-body Green's matrices, i.e.
\begin{equation}
\widetilde{G}_1 (z)=
 \frac 1{2\pi {i}}\oint_C
dz' \,\underline{g}_{x_1}(z-z')\;\underline{g}_{y_1}(z'),
 \label{contourint}
\end{equation}
where
$g_{x_1}(z)=(z-h_{x_1})^{-1}$  and
$g_{y_1}(z)=(z-h_{y_1})^{-1}$.
The contour $C$ should be taken  counterclockwise
around the continuous spectrum of $h_{y_1}$
such a way that $g_{x_1}$ is analytic on the domain encircled
by $C$. With the contour on Fig.\ \ref{contour} this mathematical condition is
met even for resonant-state energies with $z=E- \mathrm{i}\Gamma/2$.
The corresponding CS matrix elements of the two-body Green's operators in
the integrand are known exactly and analytically for all complex energies 
(see \cite{phhky} and references therein).
From this follows that all the thresholds, which correspond to the 
poles of $\underline{g}_{x_1}$, are at the right location, irrespective of
the rank $N$ used in the separable expansion.

To calculate resonances we have to find the complex zeros
of the Fredholm determinant $D(z)$ of Eq.\ (\ref{fdet}). 
Between thresholds $D(z)$ is analytic, therefore, due to a theorem of
homomorphic functions \cite{korn},
\begin{equation}
 \frac 1{2\pi {i}}\oint_{C'} D'(z)/D(z) dz = N_{C'},
\label{contour2}
\end{equation}
where $N_{C'}$ is the number of zeros inside the contour $C'$. By calculating
(\ref{contour2}) numerically we can decide whether a domain contains a
resonance or not.

\begin{figure}
\resizebox{8cm}{!}{
\includegraphics[scale=0.7]{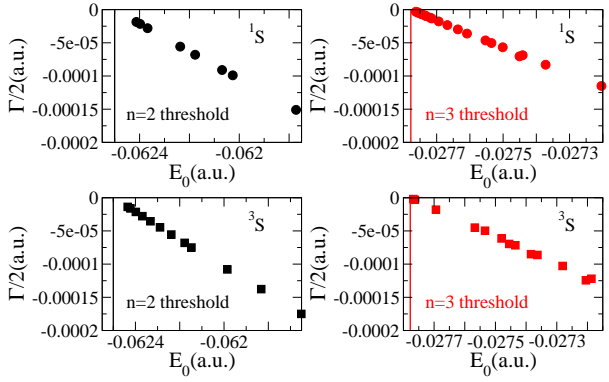}}
\caption{Accumulation of resonances above the two-body thresholds.}
\label{abra}
\end{figure}

We considered the S-wave resonances of the $e^- e^- e^+$ system.
The resonances, found at the vicinity of thresholds, 
are seen in Fig.\ \ref{abra}. The calculations were performed
with three entirely different sets of parameters: 
$x_0=18$ and $y_0=50$, $x_0=25$ and $y_0=50$, $x_0=5$ and $y_0=1000$,
while $\nu=2.1$ in all cases (the lengths are given in $a_0$ units).
We found, that the results at $N=20$ CS basis states and angular momentum
channels up to $l=\lambda=10$ are
well converged and they are rather insensitive for the 
choice of CS parameter $b$ over a broad interval.
The resonances displayed in Fig.\ \ref{abra}
are stable against the change of $x_0$ and $y_0$ parameters, 
they exhibit a remarkable $5-6$
digits stability.

We can see that the resonances are aligned along a line pointing exactly to the
two-body thresholds. As we stretched the code and went closer and closer to the
threshold we discovered more and more resonances. All of them were
along the line.
This indicates that the two-body threshold is an accumulation point of the
resonances, and probably there are infinitely many there. 

This conclusion is supported by our previous study of the  $e^+ + H$ system
\cite{huprl}, where violent oscillations of the cross sections just 
above two-body thresholds were found. Preliminary resonance calculations 
with the present method show that in the $e^+ + H$ system, where the violent
oscillations were found, there are also accumulation of resonances.

This work has been supported by the Hungarian Science Foundation (OTKA)
Grant No.\ T46791, by NSF Grant No.Phy-0243740 and by PSC
and SDSC supercomputing centers under grant No.\ MCA96N011P.

\end{document}